\documentclass[%
 reprint,
superscriptaddress,
amsmath,amssymb,
aps,
prl,
floatfix,
]{revtex4-1}
\usepackage[utf8]{inputenc}
\usepackage[danish,english]{babel}
\usepackage{amsmath}
\usepackage{amssymb}
\usepackage{graphicx}
\usepackage{color}
\usepackage[left]{lineno}
\usepackage{ulem}

\begin{document}

\newcommand{\sss}{\mathbf{s}}
\newcommand{\QQQ}{\mathbf{Q}}

\title{Distinct nature of static and dynamic magnetic stripes in cuprate superconductors}

\author{H. Jacobsen}
\affiliation{Nanoscience Center, Niels Bohr Institute, University of Copenhagen, 2100 Copenhagen Ø, Denmark}
\affiliation{Department of Physics, Oxford University, Oxford, OX1 3PU, United Kingdom}
\author{S. L. Holm}
\affiliation{Nanoscience Center, Niels Bohr Institute, University of Copenhagen, 2100 Copenhagen Ø, Denmark}
\affiliation{Interdisciplinary Nanoscience Center - INANO-Kemi, 8000 Aarhus C, Denmark}
\author{M.-E. L\u{a}c\u{a}tu\c{s}u}
\affiliation{Nanoscience Center, Niels Bohr Institute, University of Copenhagen, 2100 Copenhagen Ø, Denmark}
\affiliation{Institute of Energy Conversion, Technical University of Denmark, 4000 Roskilde, Denmark}
\author{A. T. R\o{}mer}
\affiliation{Nanoscience Center, Niels Bohr Institute, University of Copenhagen, 2100 Copenhagen Ø, Denmark}
\author{M. Bertelsen}
\affiliation{Nanoscience Center, Niels Bohr Institute, University of Copenhagen, 2100 Copenhagen Ø, Denmark}
\author{M. Boehm}
\affiliation{Institut Max Von Laue Paul Langevin, 38042 Grenoble, France}
\author{R. Toft-Petersen}
\affiliation{Helmholtz-Zentrum Berlin, 14109 Berlin, Germany}
\affiliation{Department of Physics, Technical University of Denmark, 2800 Kgs. Lyngby, Denmark}
\author{J.-C. Grivel}
\affiliation{Institute of Energy Conversion, Technical University of Denmark, 4000 Roskilde, Denmark}
\author{S. B. Emery}
\affiliation{Department of Physics and Institute of Materials Science, University of Connecticut, Connecticut 06269, USA}
\affiliation{Present address: Naval Surface Warfare Center Indian Head EOD Technology Division, Indian Head, MD 20640 USA
}
\author{L. Udby}
\affiliation{Nanoscience Center, Niels Bohr Institute, University of Copenhagen, 2100 Copenhagen Ø, Denmark}
\author{B. O. Wells}
\affiliation{Department of Physics and Institute of Materials Science, University of Connecticut, Connecticut 06269, USA}
\author{K. Lefmann}
\affiliation{Nanoscience Center, Niels Bohr Institute, University of Copenhagen, 2100 Copenhagen Ø, Denmark}

\begin{abstract}
We present detailed neutron scattering studies of the static and dynamic stripes in an optimally doped high-temperature superconductor, La$_2$CuO$_{4+y}$.
We observe that the dynamic stripes do not disperse towards the static stripes in the limit of vanishing energy transfer. Therefore, the dynamic stripes observed in neutron scattering experiments are not the Goldstone modes associated with the broken symmetry of the simultaneously observed static stripes, and the signals originate from different domains in the sample. 
These observations support real-space electronic phase separation in the crystal, where the static stripes in one phase are pinned versions of the dynamic stripes in the other, having slightly different periods.
 Our results explain 
earlier observations of unusual dispersions in underdoped 
La$_{2-x}$Sr$_x$CuO$_{4}$ ($x=0.07$) 
and La$_{2-x}$Ba$_x$CuO$_{4}$ ($x=0.095$).

\end{abstract}

\maketitle

\begin{figure*}[th!]
\includegraphics[width=1\textwidth]{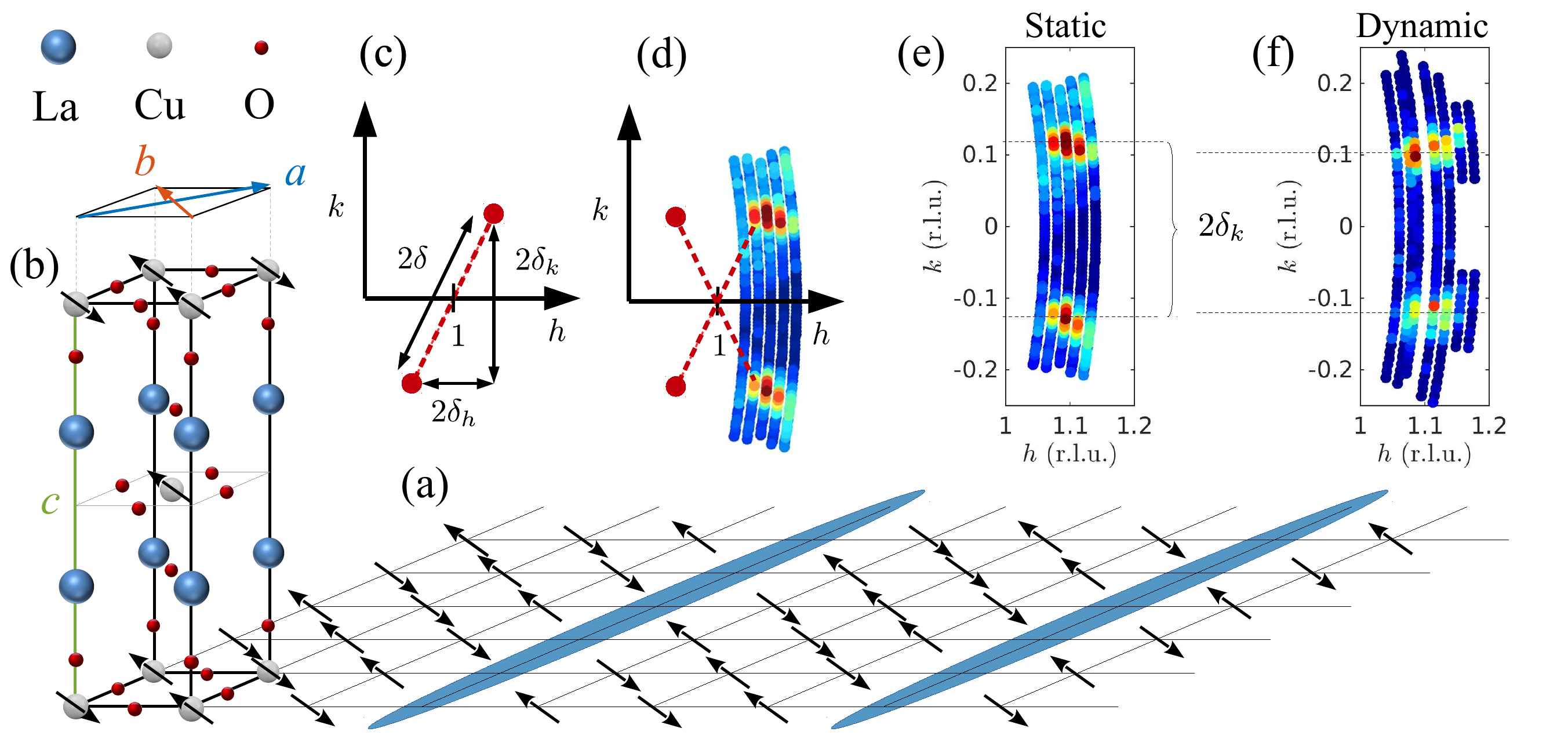}
\caption{Sketch of magnetic and charge stripes in the cuprate high-temperature superconductor La$_{2}$CuO$_{4+\delta}$ (LCO+O). 
(a) Illustration of magnetic stripes with a period of 8, concurrent with period 4 charge stripes along the Cu-O-Cu bond directions (broad blue lines). Another type of domains exists, where the stripes are rotated 90 degrees, still lying within the plane (not shown).
(b) The tetragonal unit cell of LCO+O illustrating the spins  on the Cu ions. The spins are aligned along the orthorhombic $b$-axis, shown above the unit cell.
(c) Illustration of reciprocal space (in orthorhombic notation) showing the position of the incommensurate magnetic stripe peaks for stripes approximately along the (110) direction. The difference between $\delta_h$ and $\delta_k$ is exaggerated for clarity.
(d) the quartet of peaks around the (100) position observed when stripes are present along both the (110) and (1$\bar{1}0$) directions. The coloured regions show the regions probed in the present experiment.
(e) Example of the static ($\Delta E=0$) and (f) dynamic ($\Delta E=1.5$~meV) stripe signal in LCO+O, measured by neutron scattering.
} \label{fig:illustrations}
\end{figure*}

 An imperative open question in materials physics is the nature of high-temperature superconductivity. Unlike conventional superconductors, where the Cooper pairing mechanism is well-established \cite{BCS}, the pairing mechanism in high-temperature superconductors (HTS) still sparks controversy \cite{HTSCpairs}. A comprehensive description of the electronic behavior inside HTS is indispensable to push this field of research onward. Hence, the magnetic structures which appear close to as well as inside the superconducting phase are still being studied intensively \cite{Keimer2015,Fradkin2015}. 
 In many HTS compounds, experiments indicate a modulated magnetic structure,  consistent with superconducting "stripes" of charge separated by magnetic regions
as sketched in Fig.~\ref{fig:illustrations}a \cite{Tranquada1995}. Magnetic excitations, referred to as "dynamic stripes", are found with similar periodicity, and are therefore thought to be related to  the Goldstone modes of the static stripes \cite{Vojta2009}.

Here we present evidence that this model is incomplete for a family of HTS. We find that the dynamic stripes do not disperse towards the static stripes in the limit of vanishing energy transfer and interpret this in terms of electronic phase separation, where static and dynamic stripes populate different spatial regions of the HTS.

Compounds based on the La$_2$CuO$_4$ family were the first high-temperature superconductors (HTS) to be discovered \cite{bednorzmuller86}. They become superconducting upon doping with electrons or holes, with a maximum critical temperature, $T_c\approx 40$~K, whether the dopant is Sr (La$_{2-x}$Sr$_x$CuO$_4$, LSCO), Ba (La$_{2-x}$Ba$_x$CuO$_4$, LBCO), or O (La$_2$CuO$_{4+y}$, LCO+O). 
The generic crystal structure of these compounds is illustrated in Fig.~\ref{fig:illustrations}b. They consist of planes of CuO separated by layers of La/Sr/Ba. Each Cu atom is at the center of an octahedron of oxygen atoms. 
At elevated temperatures these materials are in the high-temperature tetragonal (HTT) phase.
Upon lowering the temperature, the crystals enter the low-temperature orthorhombic phase (LTO) where the oxygen octahedra tilt around the tetragonal $a$ axes, leading to a change in lattice parameters, $a<b$ and to possible twinning \cite{Braden1992}, see Supplementary Material for details \cite{Supplementary}.

Since the first discovery, a multitude of HTS have been found in the cuprate family. The amplitude and period of the stripe order modulations vary strongly with the choice and amount of dopant, with static stripes being particularly pronounced in LCO+O \cite{Lee1999}.  

The spin stripes can be measured using magnetic neutron scattering, where they are observed as pairs of intensity peaks at incommensurate (IC) wave vector transfers, {e.g.} at ${\bf Q}=(1+\delta_h,\delta_k,0)$ and ${\bf Q}=(1-\delta_h,-\delta_k,0)$ for stripes along the (110) direction, see Fig.~\ref{fig:illustrations}c. Here, the components of the scattering vector are given in terms of  $(2\pi/a, 2\pi/b,2\pi/c)$, where $a,b$ and $c$ are the orthorhombic lattice constants. The real-space modulation period is $L \sim (2\pi /\delta) a$, and we refer to $\delta$ as the incommensurability of the stripes. Typically $\delta_h \approx \delta_k$, indicating that the modulation is approximately along the Cu-O-Cu bonds (the (110) and (1$\bar{1}0$) directions), although variations have been reported, indicating a kink in the stripes after a number of unit cells \cite{Kimura00,Lee1999}.

Typically, stripes are  observed not only at the above mentioned positions, but also at ${\bf Q}=(1-\delta_h,\delta_k,0)$ and ${\bf Q}=(1+\delta_h,-\delta_k,0)$, giving rise to a quartet of peaks around the (100) position, as illustrated in Fig.~\ref{fig:illustrations}d. This indicates that the compound exhibits stripes (approximately) along both the (110) and (1$\bar{1}$0) directions, most likely by the stripes in adjacent layers alternating between the (110) and (1$\bar{1}0$) directions \cite{Hucker2011}.

Inelastic neutron scattering has shown the presence of dynamic stripes, which at low energies have similar modulation period as the static stripes \cite{Yamadaplot}. 
The modulation period of the stripes is found to be almost constant up to energy transfers of $\Delta E \sim 10-15$~meV \cite{christensen04,lake99}. In the cuprates an hourglass shaped dispersion develops at higher energies \cite{hourglass}.

The incommensurability of the stripes varies with doping. In the LSCO-type cuprates, $\delta$ increases linearly with doping and saturates at a maximal value of $\delta =1/8$ \cite{Yamadaplot}. In some cuprates, similar stripes of charge with half the modulation period have been observed using X-ray diffraction, validating the picture of magnetic and charge stripes in Fig.~\ref{fig:illustrations}a \cite{YBCOchargestripes, Silva2014, Thampy2014, Croft2014,Chen2016}. However, the energy resolution of X-rays does not allow to distinguish between static and dynamic stripes.

We have used elastic and inelastic scattering of low energy neutrons to accurately measure the reciprocal space position of the static and dynamic stripes in highly oxygenated LCO+O in the LTO phase.  The experiments were   performed at the cold-neutron triple axis spectrometers FLEXX at HZB, Berlin \cite{FLEXX}, and ThALES at ILL, Grenoble \cite{ThALES_data}. The elastic energy resolution in the ThALES experiment was 0.24 meV (Full Width at Half Max, FWHM), while the $Q$ resolution was 0.05 r.l.u. (FWHM).
For further details on the experiments, see the Supplementary Material \cite{Supplementary}.

\begin{figure}[t]
\includegraphics[width=0.48\textwidth]{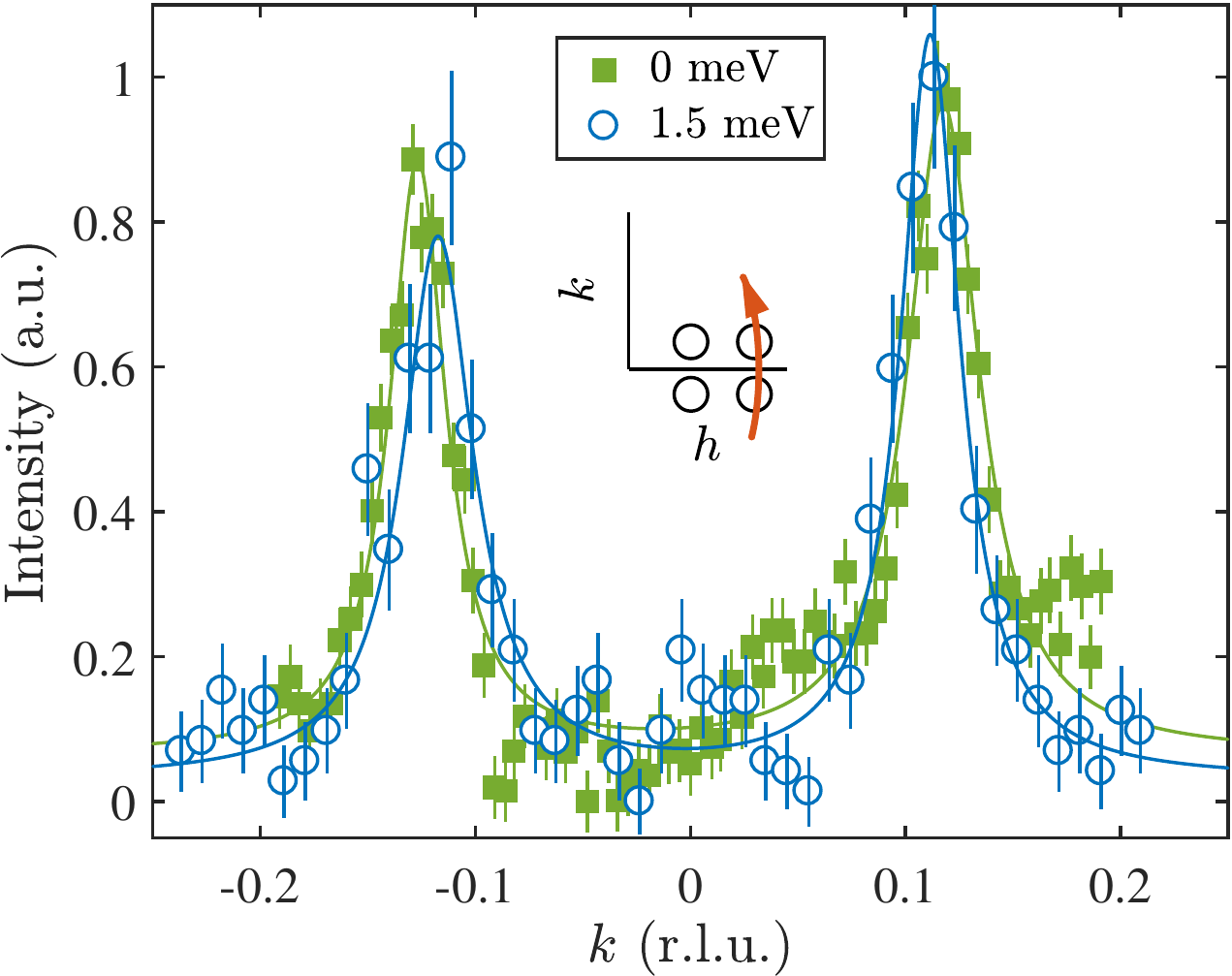}
\caption{Neutron scattering data for LCO+O scanned along the direction shown in the insets, showing the shift in peak position between the elastic stripes (green) and low-energy inelastic stripes (blue). The data have been rescaled and the background subtracted.} \label{fig:rawdata}
\end{figure}

Panel (d) of Fig.~\ref{fig:illustrations} shows how we probe two of the four IC peaks in our neutron scattering experiments. The actual data for a series of scans are shown in panels (e) and (f) as 2D colorplots. Fig.~\ref{fig:rawdata} shows examples of the scans through the center of the peaks at 0 and 1.5~meV energy transfer, probing the static and dynamic stripes, respectively. The inset illustrates the direction of the scans in reciprocal space.

\begin{figure}[t]
\includegraphics[width=0.48\textwidth]{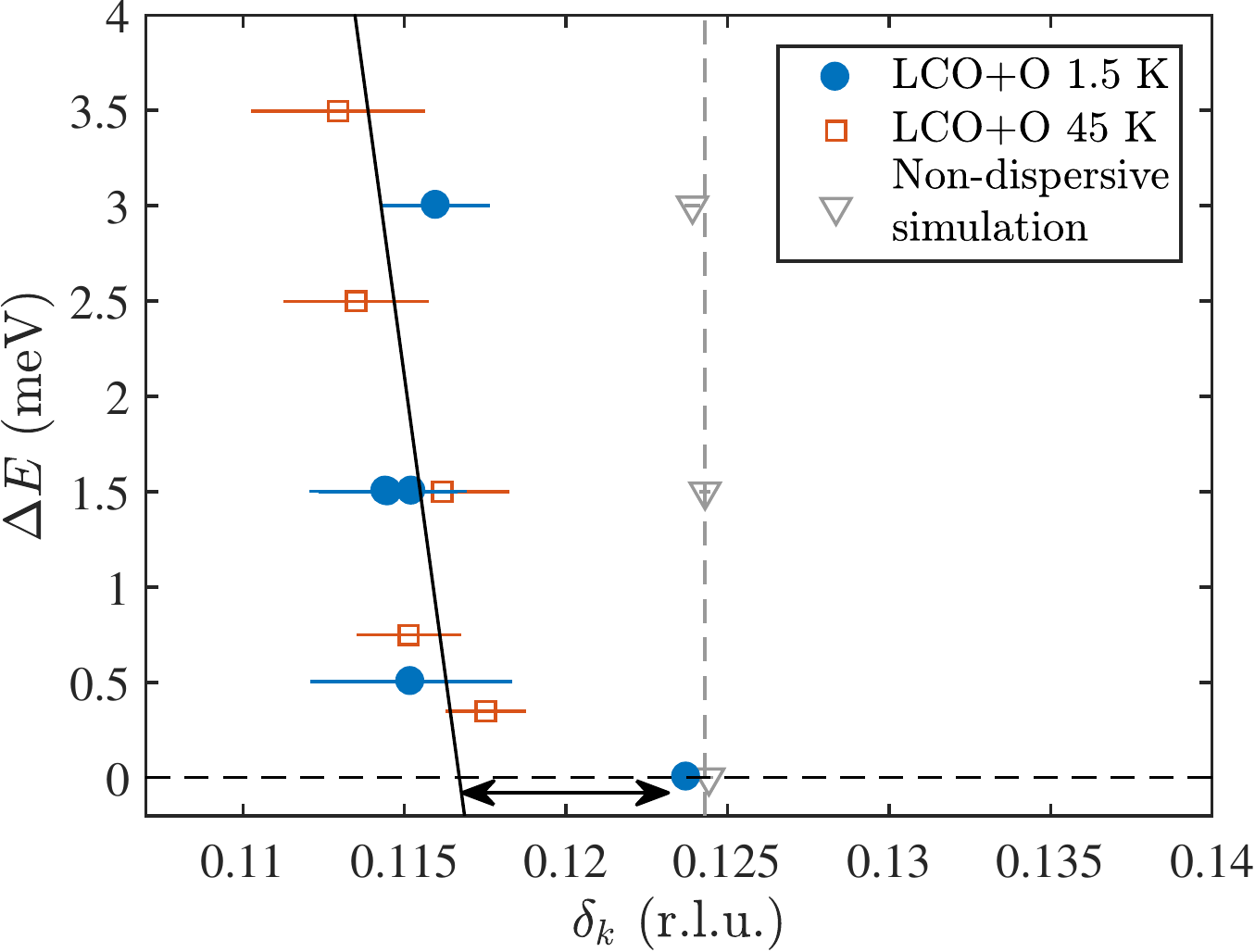}
\caption{ The incommensurability, $\delta_k$ at different energy transfers $\Delta E$ for LCO+O. A significant shift is seen between the elastic and inelastic data. The solid black line is a linear fit to the dispersion for $\Delta E>0$. Gray triangles represent the dispersion relation obtained from simulated data, where the simulated dispersion relation is vertical. 
}
\label{fig:dispersion}
\end{figure}

 To eliminate errors from minor misalignments, we determine the incommensurability along $k$, $\delta_k$, as half the distance between the peak centers.  In Fig.~\ref{fig:dispersion} we display $\delta_k$  for all energy transfers probed in the experiment at two temperatures. As expected, the dynamic stripes appear at the same reciprocal space position in the normal phase (45~K) as in the SC phase (2~K) (within the instrument resolution), whereas the static stripes are only present at low temperature. The elastic stripes are found to be rotated by $7^\circ$ from the Cu-O-Cu bond directions, while the observed inelastic stripes are rotated by $3^\circ$.
The inelastic dispersion appears continuous and steep, consistent with earlier cuprate results \cite{christensen04,lake99}. However, the elastic signal shows a large and significant difference in $\delta_k$, appearing as a discontinuity in the dispersion relation at vanishing energy transfer. 
Similar observations have been briefly remarked upon in underdoped 
La$_{2-x}$Sr$_x$CuO$_{4}$ ($x=0.07$) \cite{Jacobsen2015} and 
La$_{2-x}$Ba$_x$CuO$_{4}$ ($x=0.095$) \cite{Xu2014}. In both cases the observation was left unexplained.

To rule out that these surprising differences in $\delta_k$ and the stripe rotation are artifacts caused by experimental non-idealities, we have performed a virtual ray-tracing experiment using a close model of our experiment, further detailed in the Supplementary Material \cite{Supplementary}. This method is known to accurately reproduce experimental effects like peak broadening and displacement  \cite{udby11}. The virtual experiments exclude misalignment of the instrument as a cause of the effect and show that the experimental resolution can cause a tiny shift in the observed incommensurability, see Fig.~\ref{fig:dispersion}. 

The experimentally observed shift in peak position is, however, more than an order of magnitude larger than what can be explained by instrument effects, and is therefore a genuine property of the sample. Hence, in order not to violate the Goldstone's theorem, the static and dynamic stripes must originate from different regions in the sample. There are two probable ways this can occur:

First, the dynamic stripes could be transverse fluctuations from the static stripe order, resembling ordinary spin waves. 
Due to the neutron scattering selection rules, the scattering observed in the elastic and inelastic channels stem from different twin domains as explained in detail in the Supplementary Material \cite{Supplementary}. This results in a shift in the observed peak position between the elastic and inelastic channels, comparable to the observed shift. The magnitude and direction of the shift due to twinning depends heavily on 
$\delta_h$ and $\delta_k$ and requires $\delta_h<\delta_k$. 
Secondly, the static and dynamic spin response may originate from different microscopic regions which are not related by twinning. This suggests a real-space electronic phase separation of the crystal into regions with two different spin structures; one domain type which has static stripe order and associated dynamic stripes, and another type of domain where only dynamic stripes are present.

At first glance the twinning model seems to provide an explanation of our data. However, it fails to explain the similar observations in LBCO (where $\delta_h=\delta_k$) mentioned above {\cite{Xu2014}}, as the model requires $\delta_h<\delta_k$. Furthermore, the model relies on the assumption that the four twin domains display only one type of stripe order with associated transverse excitations. 
Most likely these assumptions are too simplified and
relaxing any of them  reduces the effect of twinning on the observed signal. We therefore turn to the second model: electronic phase separation.

Muon spin rotation experiments on highly oxygenated 
LCO+O show that the material electronically phase separates into a magnetic (A) and a superconducting (B) phase of roughly equal volume  \cite{LCOOmuonpaper,Mohottala2006} and transition temperature $T_c\approx T_N\sim 40$ K with the present slow cooling conditions. Based on these experiments we propose the following properties of the two phases:

Phase A is underdoped (resembling LSCO with $n_h=0.125$) and has static magnetism (and weak fluctuations), responsible for the observed static signal and a small fraction of the dynamic signal. Phase B is optimally doped (resembling LSCO with $n_h=0.16$) and superconducting with strong fluctuations, responsible for (the majority of) the observed dynamic signal. 

We note that no spin gap was observed below $T_c$ in our experiments. Absence of a spin gap was also observed in the experiments on strongly underdoped LSCO \cite{Jacobsen2015} and LBCO \cite{Xu2014}, mentioned above. Both materials were suggested not to be d-wave superconductors but instead display Pair Density Wave (PDW) superconductivity \cite{Fradkin2015,Christensen2016}. Our results are consistent with this interpretation. A PDW state would require some degree of magnetic order in the SC phase, but this may be extremely weak and thus effectively invisible in our experiments. 
The simultaneous observation of gapless excitations and a shift in incommensurability in all three compounds suggests a connection between the two effects. The gapless excitations are likely a result of a PDW state in the sample, while the shift is caused by electronic phase separation. At present it is unclear whether these two behaviours are related. 


The critical temperature of the superconducting phase, $T_c$, coincides with the N\'eel temperature of the magnetic phase, $T_N$, such that above this temperature superconductivity and the static magnetism disappear, but strong stripe fluctuations remain. The fact that $T_c\approx T_N$ is likely not coincidental, but it is unclear whether the electronic phase separation is caused by, or is the cause of, the close proximity of $T_c$ and $T_N$. We suggest a scenario where the ground state energy for phase A and phase B are very close and lowering the temperature below $T_c$ will cause an electronic phase separation with concurrent static magnetism and superconductivity. 
The spatial distribution of impurity potentials as well as inhomogeneous hole doping becomes important parameters that can tip a region towards becoming type A or B, e.g. by pinning.


The relative population of each phase is primarily controlled by the total number of holes, but can also be influenced by an applied magnetic field or by crash cooling \cite{romer15,Lee1999}.
In the case of LCO+O, crash cooling can further inflict a lowering of $T_c$ which has been explained by disconnection of the optimally superconducting pathways \cite{fratini2010}.

The crucial point is that although the two phases are closely related, there is no a priori reason why the stripe order in phase A and the stripe dynamics in phase B should have the same incommensurability. Indeed, our results show that this is not the case. This indicates that other properties of the stripes may not be identical either, and one should thus be extremely careful when interpreting neutron scattering data on stripes.

Phase separation has been suggested to occur in a number of cuprates or related compounds, a few of which we will mention here. LSCO with $x=0.12$  has been suggested to phase separate into microscopic superconducting regions with gapped dynamic stripes and non-superconducting regions with static stripes \cite{Kofu2009}. Spontaneous, microscopic  phase separation has also been observed in purely oxygen doped LCO+O crystals \cite{Mohottala2006} and in crystals doped with both oxygen and strontium \cite{Udby2013}. Furthermore, recent studies of La$_{5/3}$Sr$_{1/3}$CoO$_4$ show evidence of microscopic phase separation into components with different local hole concentration \cite{Drees2014,Babkevich2016}.
In the latter material the upper and lower parts of the hourglass dispersion  are even proposed to originate from different nano-scale structures in the sample \cite{Drees2014}. No discrepancy in the incommensurability between static and dynamic stripes was reported in these studies.

The idea of dynamic and static stripes having different origin is supported by a number of other observations. 
For example, the static and dynamic stripes exhibit different behaviors as function of temperature. In underdoped LSCO and in LCO+O as evidenced in this experiment, the static stripes vanish above $T_c$, but the dynamic stripes remain to far higher temperatures \cite{Lee2000,lake02,lake01}. In contrast, in the optimally doped region, the static stripes are altogether absent, while the dynamic stribe exist above a certain energy gap \cite{lake99}. In the heavily overdoped region it has been shown that substituting small amounts of Fe for Cu induces static magnetism \cite{He2011}. The incommensurability of the induced magnetic order is governed by nesting of the underlying Fermi surface and differs from the $1/8$ periodicity of the low-energy dynamic stripes.

When applying a magnetic field, the static stripes are in general strengthened \cite{lake01,lake02,chang09,romer13,chang08,leePRB69}, with a few exceptions \cite{Udby2009,chang08}. In many cases this happens with an accompanying change in the dynamic stripe spectrum \cite{lake01,lake02,chang09}, but in other cases, the dynamic stripe spectrum is unchanged \cite{romer13}. 
Hence, the coupling between static and dynamic stripes is not simple and unique. 

In conclusion we have found that the dynamic stripes do not disperse towards the static stripes in the limit of vanishing energy transfer in a HTSC. The effect is subtle and requires high flux and good resolution such as provided by the ThALES spectrometer in order to be observed. Our findings are, however, of prime importance, since they suggest that the observed static and dynamic stripes originate from different electronic phases in the sample, where one of these phases is likely to be a competitor for superconductivity with the development of static stripe order.  

Our observations are relevant for all compounds displaying stripe order. As an example, the structurally similar, but non-superconducting compound (La,Sr)$_2$NiO$_4$ (LSNO) displays magnetic and charge stripes with the dynamic stripes persisting at higher temperatures than the ordering temperature \cite{Anissimova2014}. In some of these compounds it has also been  observed that the ordering vector of the static and dynamic stripes do not coincide at vanishing energy transfer \cite{Freeman_personal}. We speculate that a similar electronic phase separation could be in play here, as we suggest for LCO+O. It is likely that this mechanism also explains earlier observations of unusual dispersions in LSCO \cite{Jacobsen2015} and LBCO \cite{Xu2014}. Our findings may thus be a vital part in unveiling the nature of high temperature superconductivity.
 

\begin{acknowledgments}
We thank ILL, Grenoble, France, and HZB, Berlin, Germany for providing us access to their neutron scattering facilities. 
We are indebted to E. Farhi for providing us with a model of the ILL neutron guide system for use in the Monte Carlo simulations. We thank N.\ B.\ Christensen, P. J. Ray, J. M. Tranquada, J.I. Budnick, P.G. Freeman, M. Skoulatos, and D. Prabhakaran for illuminating discussions. We thank P. J. Ray for help with some of the figures.

Work at University of Connecticut was supported by the USDOE Basic Energy Sciences under contract DE-FG02-00ER45801. 
The work was supported by the Danish Research Council FNU through the grants DanScatt and Magnetism in Superconductors.
\end{acknowledgments}

\clearpage
\appendix

\section{Sample details}
The samples were prepared by growing stoichiometric LCO crystals at the Technical University of Denmark in an optical image furnace using the travelling solvent float zone technique. After annealing and characterization, chosen crystals were super-oxygenated in an aqueous bath at the University of Connecticut. The resulting LCO+O crystals were cut in pieces suitable for neutron scattering experiments and smaller pieces of the sample were characterized by resistivity and susceptibility measurements to find a superconducting transition temperature of $40 \pm 1$~K, typical for LCO+O .

In equilibrium LCO+O spontaneously separates into a series of superconducting and magnetic line phases. If out of equilibrium due to gross oxygen inhomogeneity, crystals often contain both $T_c=30$~K and $T_c = 40$~K superconducting regions \cite{Mohottala2006}.

\begin{figure}
    \centering
    \includegraphics[width=0.48\textwidth]{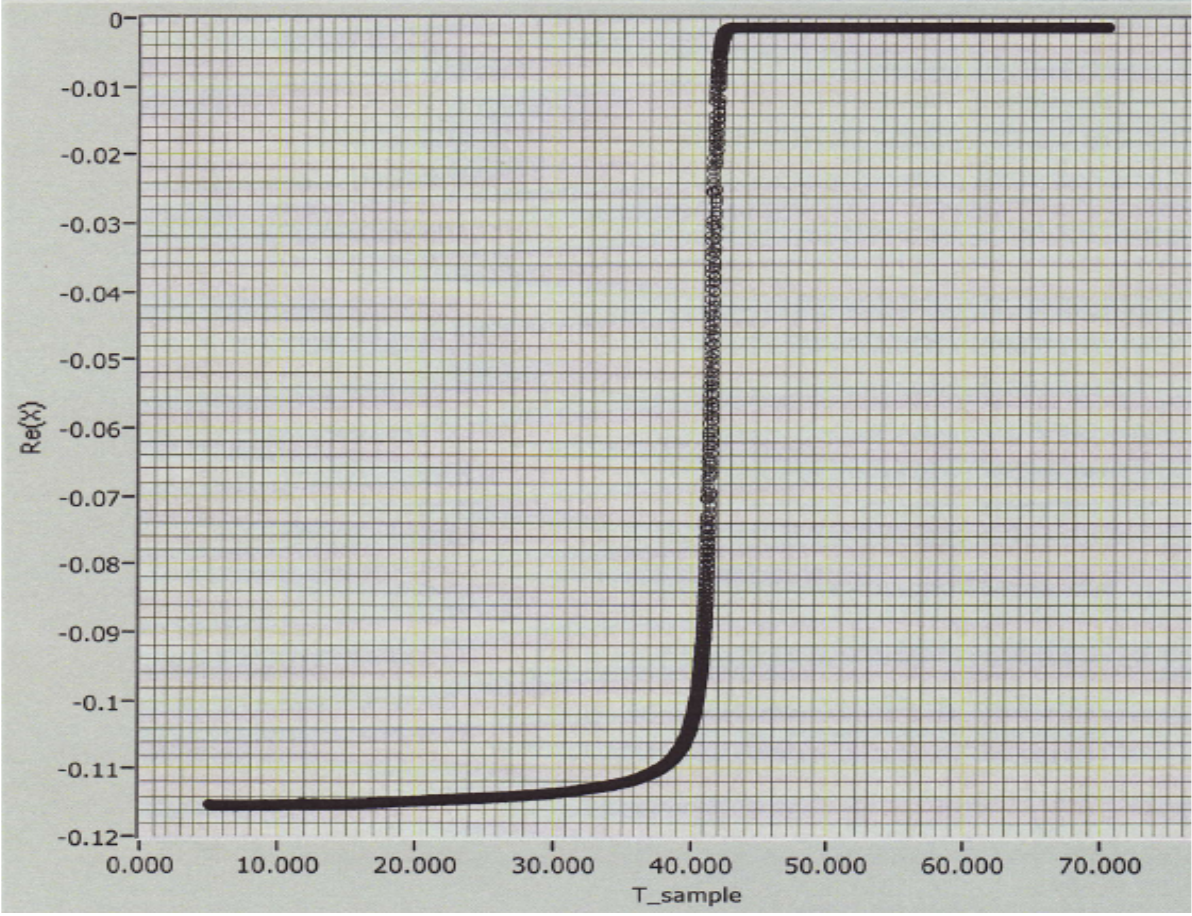}
    \caption{The susceptibility of the LCO+O sample in a weak field, showing the SC transition clearly around $T_C\sim 40$~K. }
    \label{fig:susceptibility}
\end{figure}

The superconducting transition temperature for our sample was measured by a vibrating sample magnetometer. The data is shown in Fig.~\ref{fig:susceptibility} and show that our sample clearly has only one superconducting transition near $T=40$~K, and therefore just one superconducting phase that is similar to what was obtained by Lee {\em et al.} 

\cite{Lee1999}.
The orthorhombic lattice parameters are $a=5.33$~\AA{}, $b=5.40$~\AA{}, $c=13.2$~\AA{}.
  The spins align along the $b$ axis in LCO+O, just as for the parent compound \cite{Lee1999}.

\section{Details on neutron scattering experiments on LCO+O}
Neutron scattering experiments were   performed at the cold-neutron triple axis spectrometers FLEXX at HZB, Berlin \cite{FLEXX}, and ThALES at ILL, Grenoble \cite{ThALES_data}.
The spectrometers were configured to run at a constant final energy of $5.0$~meV. 

At FLEXX the sample was mounted inside a magnet. The results of applying a magnetic field will be reported in a following publication.
In both experiments, a velocity selector in the incident beam before the monochromator removed second-order contamination, while a
cooled Be-filter between sample and analyzer further reduced background. The sample was aligned in the $(a,b)$ plane. At FLEXX, several cylindrical crystals were co-aligned, resulting in a total sample mass of $\sim 9$~g. At ThALES, only the largest crystal of mass 3.44~g was used in order to improve $Q$ resolution. 
In both experiments, we used vertically focusing monochromators, leading to relatively loose vertical collimation along the $c$-direction, where the stripe signal from cuprates is nearly constant \cite{romer15}. At the ThALES experiment there was a small offset in the A4 angle which was corrected for in the subsequent data analysis.

\begin{figure}
\centering
\includegraphics[width=0.48\textwidth]{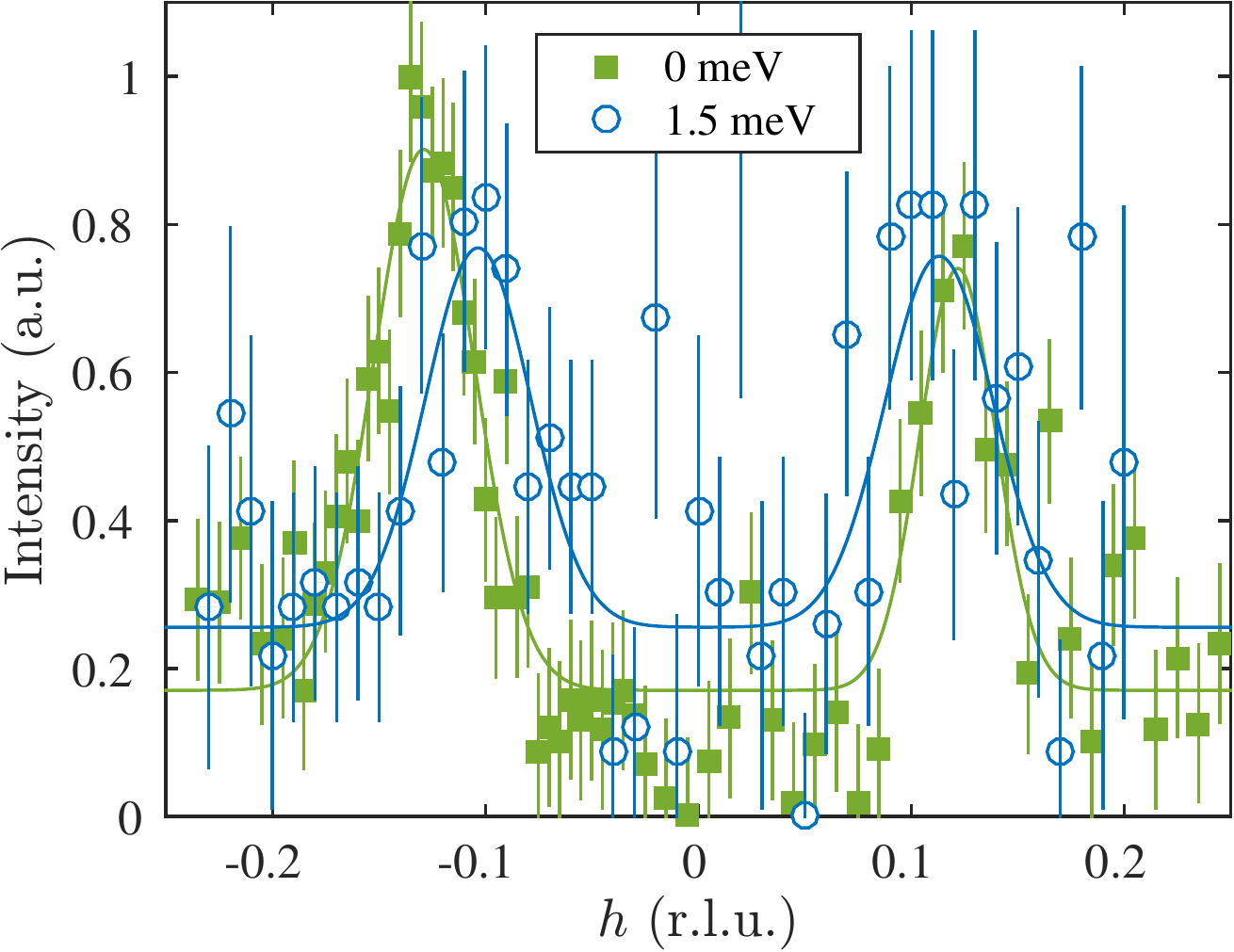}
\caption{Background subtracted  neutron  scattering  data on LCO+O, measured at FLEXX, HZB. The shift of the peak position between elastic and inelastic data is also seen here.}
\label{fig:FLEXX_data}
\end{figure}

\begin{table*}
\begin{center}
\begin{tabular}{|l|l|l|l|l|l|l|l|}\hline
& &  \multicolumn{2}{c|}{Peak 1}  & \multicolumn{2}{c|}{Peak 2}  &\multicolumn{2}{c|}{Incommensurability}   \\\hline
& Position & $h$ & $k$ & $h$ & $k$ & $\delta_h$ & $\delta_k$\\\hline
Elastic & (100) & 1.0965(6) & 0.1151(4) & 1.0969(7) & -0.1324(5) & 0.0967(5) & 0.1237(3)\\\hline
Inelastic & (100) & 1.1013(20) & 0.1058(11) &   1.1063(21) & -0.1236(13) & 0.1038(15) & 0.1147(8)\\\hline
Elastic & (010) & 0.1249(5) & 1.0944(7) & -0.1216(4) & 1.0956(7)) & 0.1233(3) & 0.0950(5)\\\hline
\end{tabular}
\end{center}
\caption{The fitted positions of the two-dimensional gaussian peaks given in r.l.u.}
\label{tab:peak_pos}
\end{table*}
The effect reported in the main paper was also seen in the experiment at FLEXX. An example of the data is shown in Fig.~\ref{fig:FLEXX_data}, showing the same difference between elastic and inelastic stripe positions as found at the ThALES experiment. In most of the experiment, however, only a single peak was measured due to time constraints.
 
\begin{figure}
\centering
\includegraphics[width=0.48\textwidth]{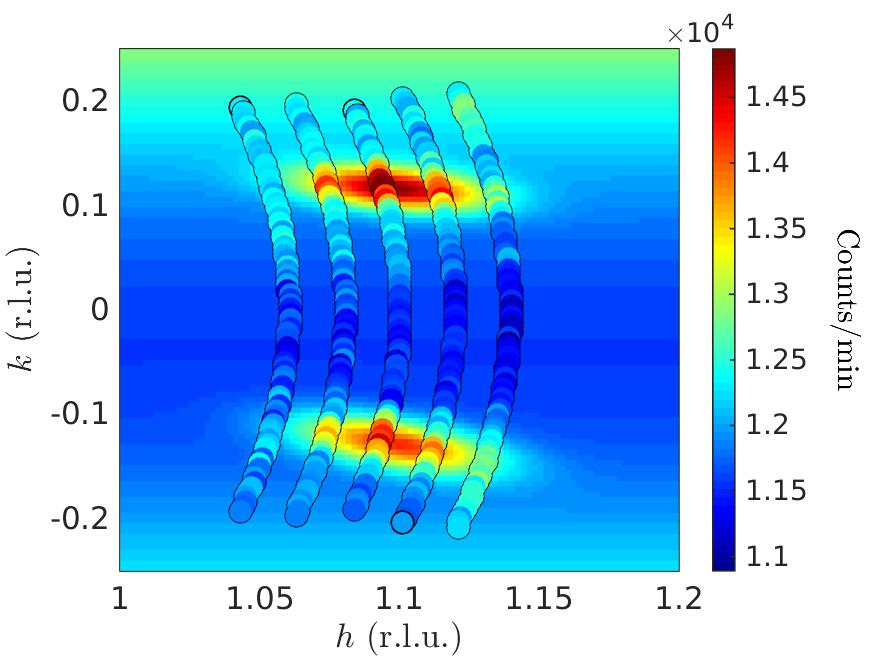}
\includegraphics[width=0.48\textwidth]{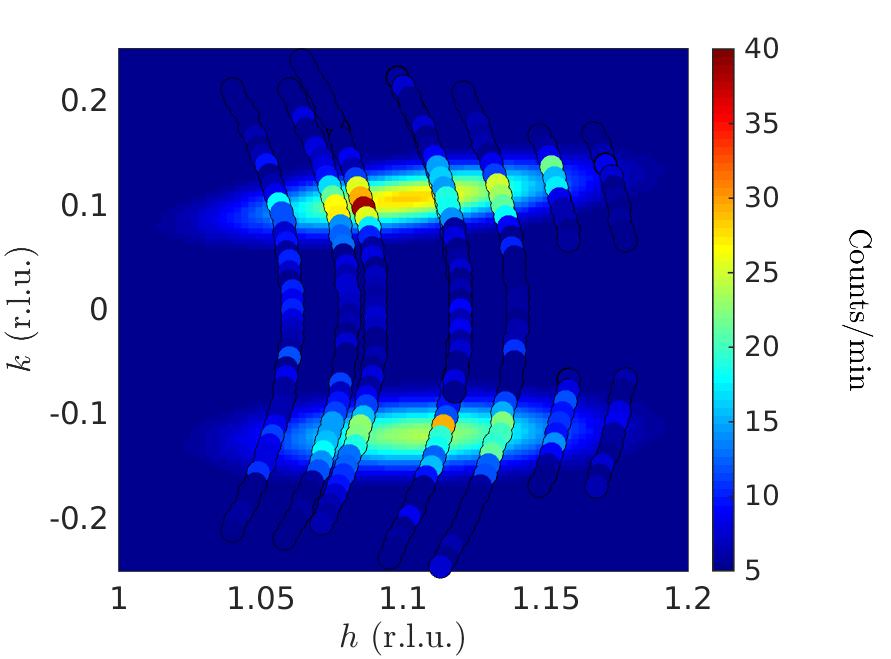}
\caption{Two-dimensional fits of the LCO+O data at $\Delta E=0$~meV (top) and 1.5~meV (bottom). The colored circles show the data, while the area around the circles  shows the fit as described in the text.}
\label{fig:supplementary:2d_fits}
\end{figure}

We here show additional data from the ThALES experiment. The measurements at $\Delta E= 0$~meV and 1.5~meV were taken as grid scans in the $(h,k)$ plane around the (100) position. The data were fitted to a pair of two-dimensional Gaussians, as seen in Fig.~\ref{fig:supplementary:2d_fits}. The fitted peak positions are given in Tab.~\ref{tab:peak_pos}

From the fits, the shift in the incommensurability between $\Delta E=0$~meV and 1.5~meV is $(-0.0070(15),0.0090(9),0)$.

\begin{figure}
\centering
\includegraphics[width=0.48\textwidth]{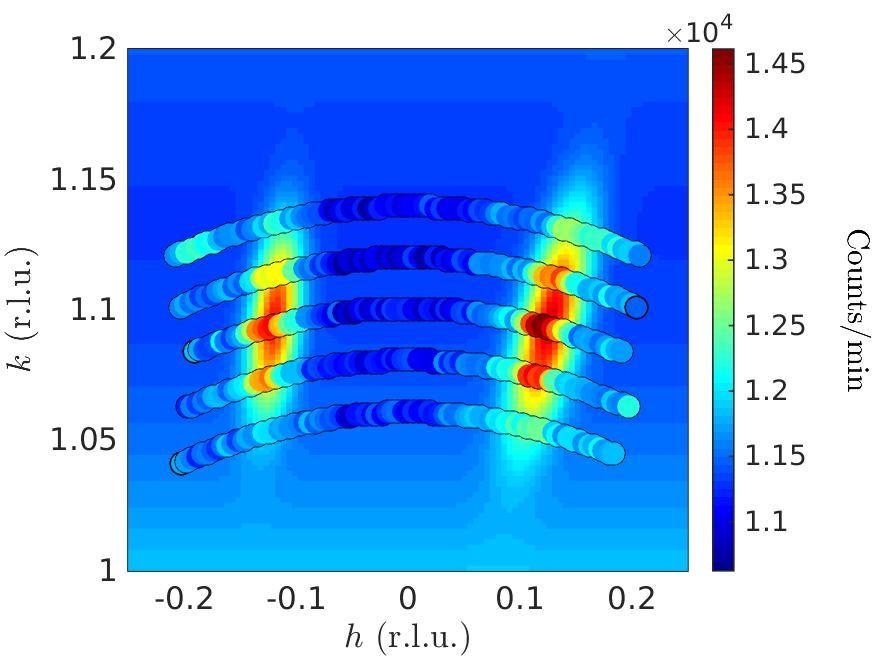}
\caption{Two-dimensional fit of the  LCO+O data at 0~meV near the (010) position. The colored circles show the data, while the area around the circles shows the fit as described in the text.}
\label{fig:supplementary:2d_fits_010}
\end{figure}

The elastic data around the (010) peak, similar to the data around (100) except for a 90 degree rotation are shown in Fig.~\ref{fig:supplementary:2d_fits_010}. The incommensurability is the same within error bars for the two data sets.
The  static stripes are rotated by approximately $7^\circ$ away from the Cu-O-Cu directions; a value that is approximately twice what was found earlier \cite{Lee1999}.

\begin{figure}
\centering
\includegraphics[width=0.48\textwidth]{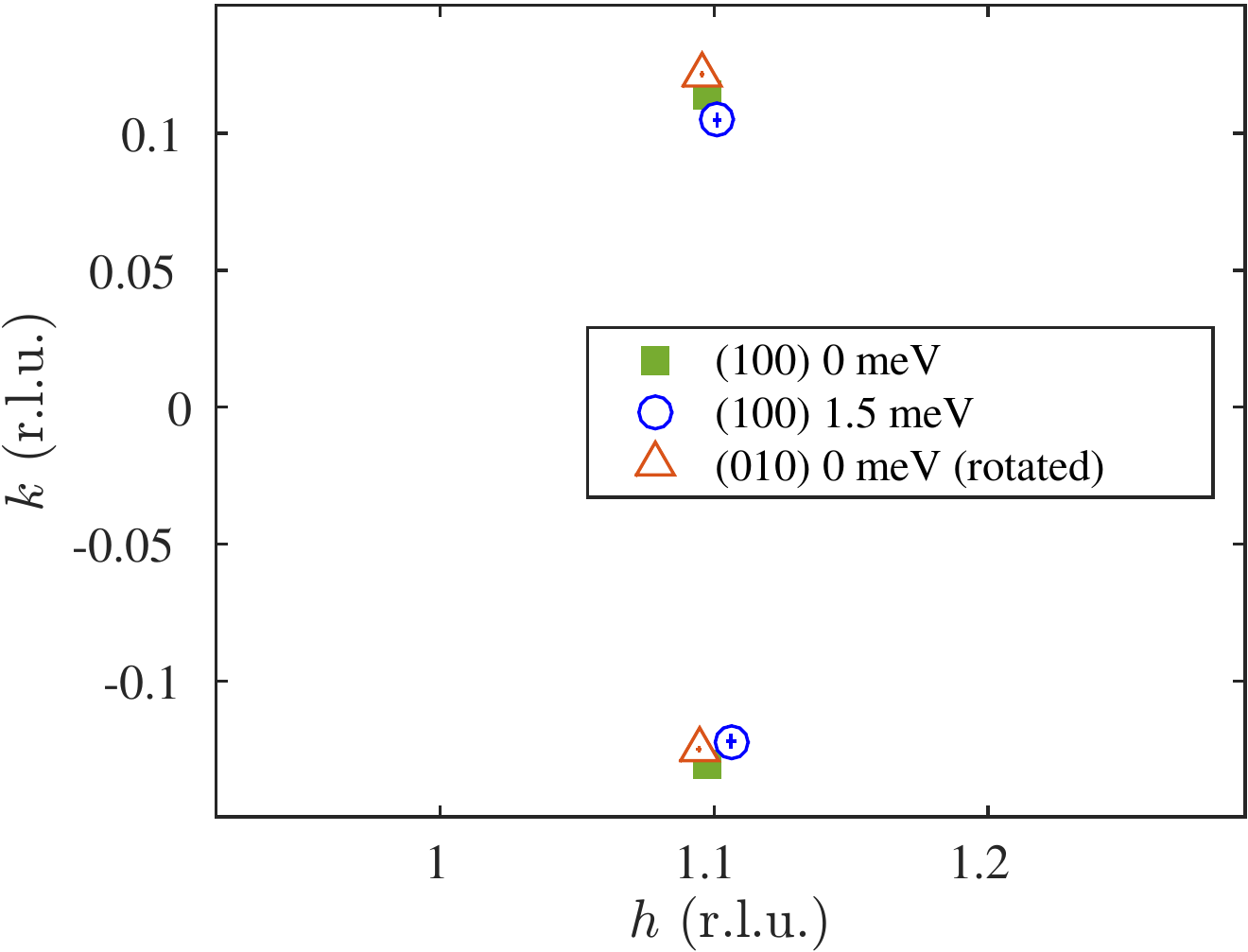}
\caption{The fitted positions of the two-dimensional gaussian peaks, showing how the inelastic peaks move toward smaller values of  $|k|$ in  LCO+O.}
\label{fig:fitted_peak_positions_2d}
\end{figure}

The peak positions of the fitted two-dimensional Gaussians are shown in Fig.~\ref{fig:fitted_peak_positions_2d}. The peak positions from the (010) peak were rotated 90 degrees for comparison. In this figure, it is observed that the inelastic peaks move together in $k$ and to larger $h$, compared to the elastic peaks. 

\begin{figure}[h]
\includegraphics[width=0.48\textwidth]{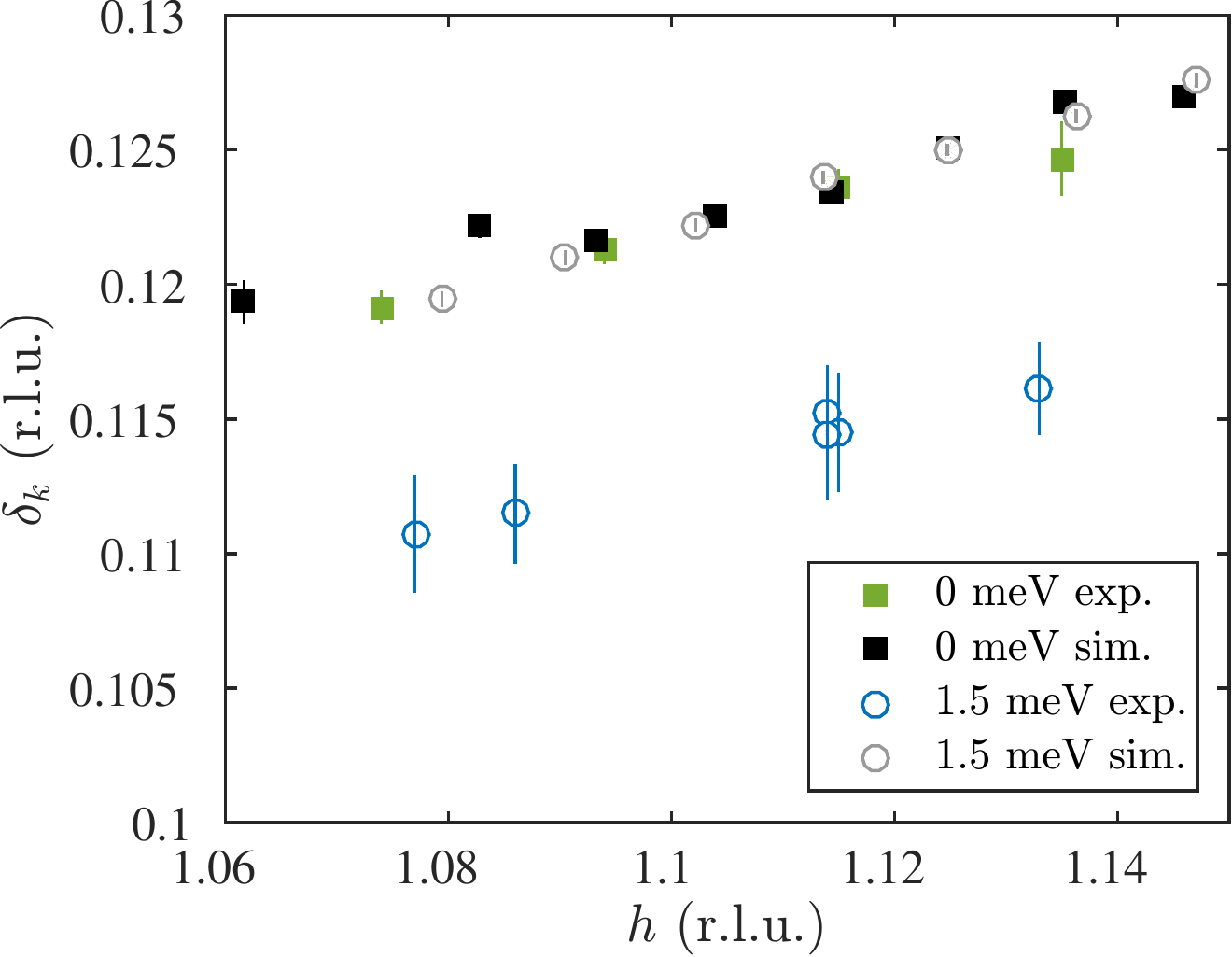}
\includegraphics[width=0.48\textwidth]{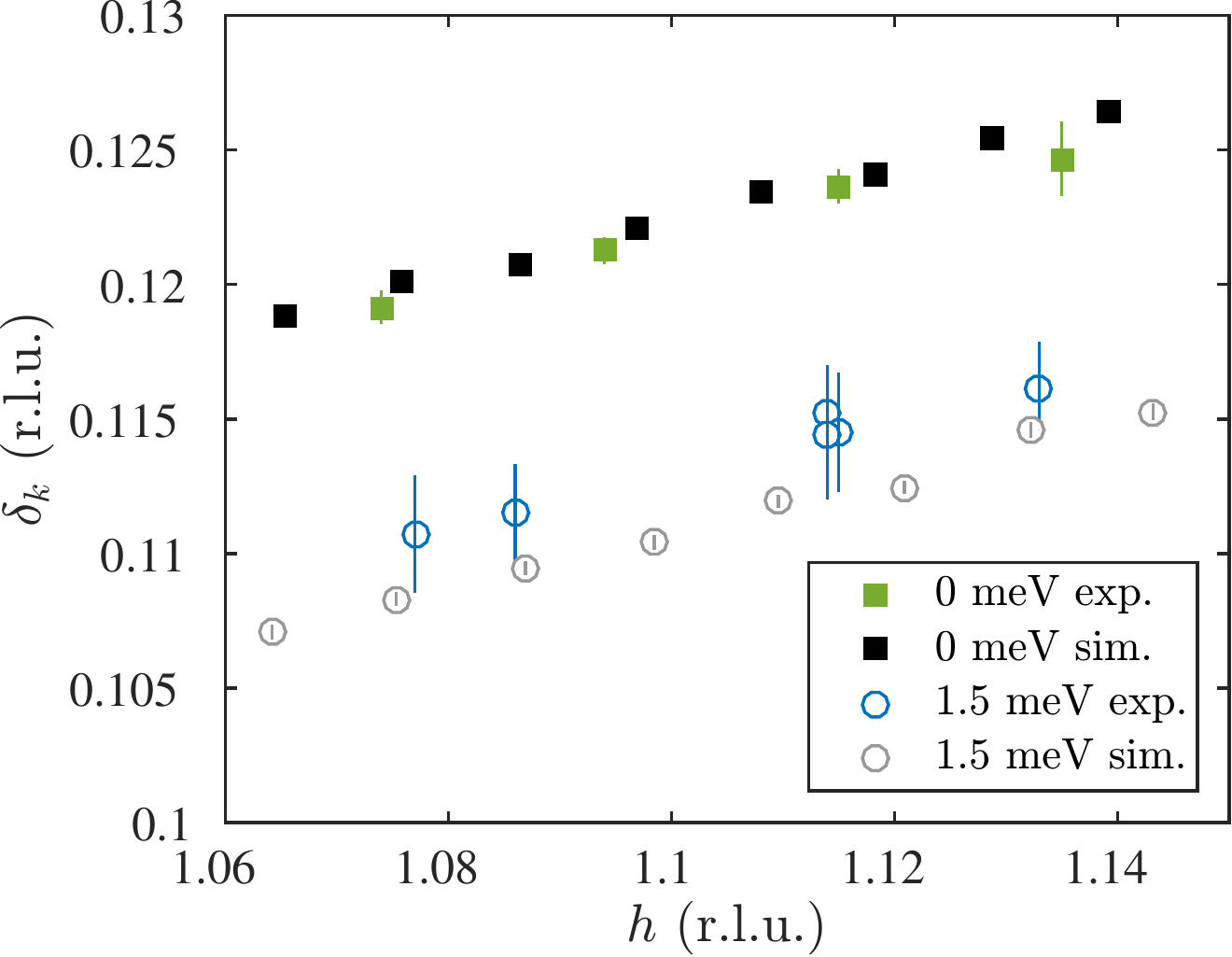}
\caption{Incommensurability along $k$ as function of $h$ for LCO+O at $\Delta E=0$~meV and 1.5~meV, along with simulated data. Top: Measured data compared to simulations where $\delta_h=\delta_k = 1/8$. Bottom: Measured data compared to simulations where $\delta_h$ and $\delta_k$ are the same as found from the experiments as detailed in the text. }
\label{fig:delta_vs_h}
\end{figure}

The data show that the distance between the incommensurate peaks is dependent on where exactly on the peak the measurement is performed. 
The shape of the resolution function and the peaks implies that the observed value of $\delta_k$ increases as function of $h$. This could lead to systematic errors, if different parts of the peak were probed at different energies. By making the grid scans shown in Figs.~\ref{fig:supplementary:2d_fits} and \ref{fig:supplementary:2d_fits_010}, we measured exactly this effect. Each scan in the grid was fitted individually. The resulting value of $\delta_k$, determined as a function of $h$, is shown in Fig.~\ref{fig:delta_vs_h}. It is seen that $\delta_k$ increases with increasing $h$, as we move through the peak. However, it is also evident that at any given position in $h$, a significant shift in $\delta_k$ happens between the elastic and inelastic signals. 

\section{Virtual experiments}

The full neutron scattering experiment at ThALES was simulated using the Monte Carlo (MC) ray tracing program McStas \cite{McStas1,McStas2}, which has previously been shown to produce very accurate results regarding, in particular, instrument resolution \cite{udby11}. Here follows a more detailed description of the simulation method.

The guide system at ILL has been simulated by E. Farhi \cite{farhi15}, and we adopted his McStas model. The remainder of the instrument was simulated using standard components from McStas: slits, graphite crystals, and a detector. For the purpose of these simulations, a  sample model was written, simulating scattering from static and dynamic stripes. This sample scatters elastically at a user-defined position in $(h,k)$-space (with the scattering being independent of $l$, as is true to a good approximation near $l=0$ where the experiment was performed \cite{romer15}). Furthermore, the sample scatters inelastically at a (possibly different) user-defined position in $(h,k)$-space. The absolute scattering cross section of the elastic and inelastic scattering can be set individually. For simplicity, and since we are not interested in absolute intensities, the cross section was kept independent of energy transfer, and no incoherent background was simulated.
A combination of two such samples, rotated with respect to each other, was used to simulate the two measured peaks at $q=(1+\delta_h,\pm \delta_k,0)$.

A small offset in the A4 angle in the experiment caused the lattice parameters to appear slightly larger than their actual values. This was accounted for in the simulations. 

Two sets of simulations were made: one with both the elastic and inelastic peaks at $\delta_h=\delta_k = 1/8$, and one with $\delta_h= 0.0973$, $\delta_k=0.1222$ for the elastic peaks and $\delta_h=0.1036$, $\delta_k=0.1133$ for the inelastic peaks, as found in the experiments.

\begin{figure*}[ht]
\centering
\includegraphics[width=0.48\textwidth]{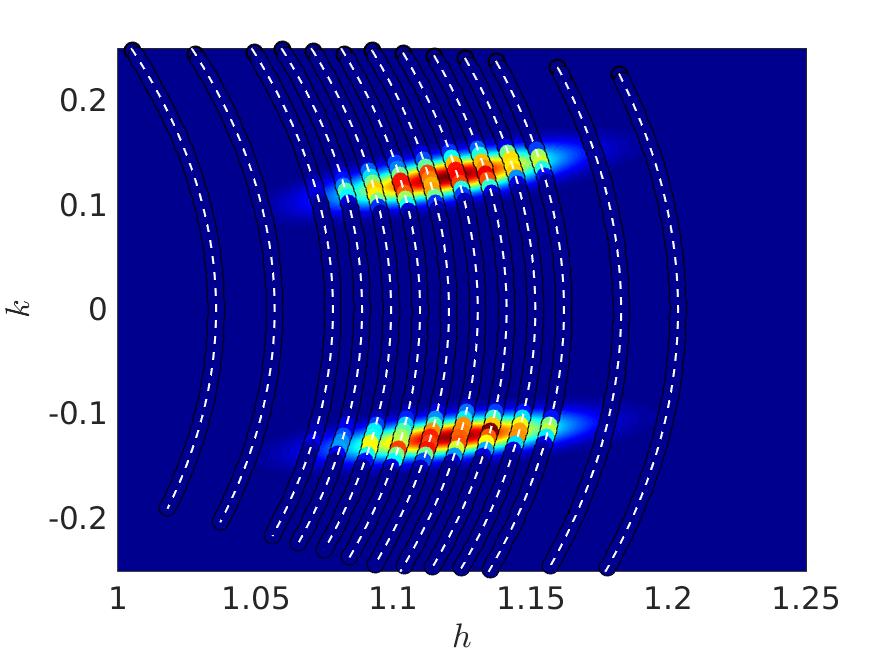}
\includegraphics[width=0.48\textwidth]{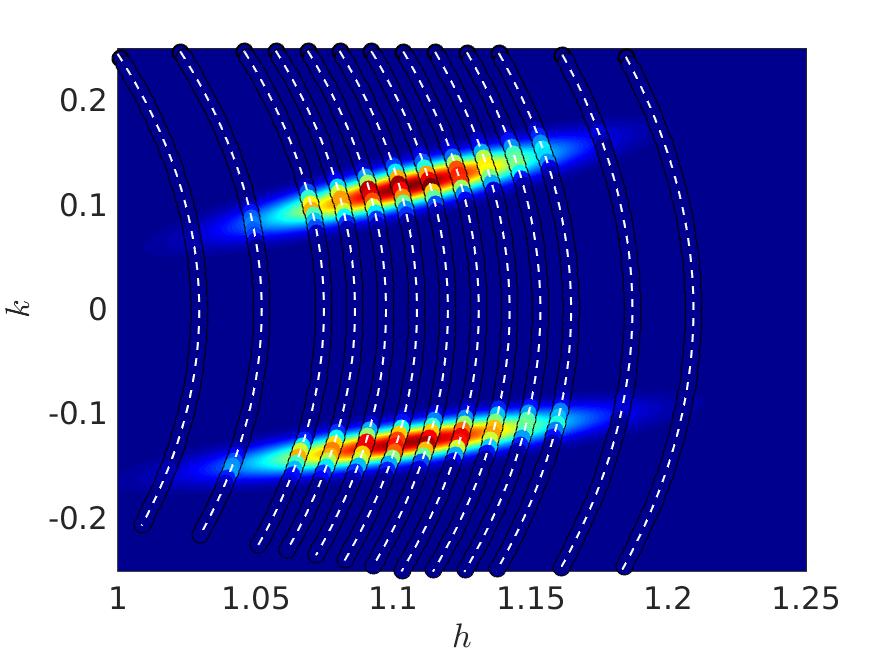} \\
\includegraphics[width=0.48\textwidth]{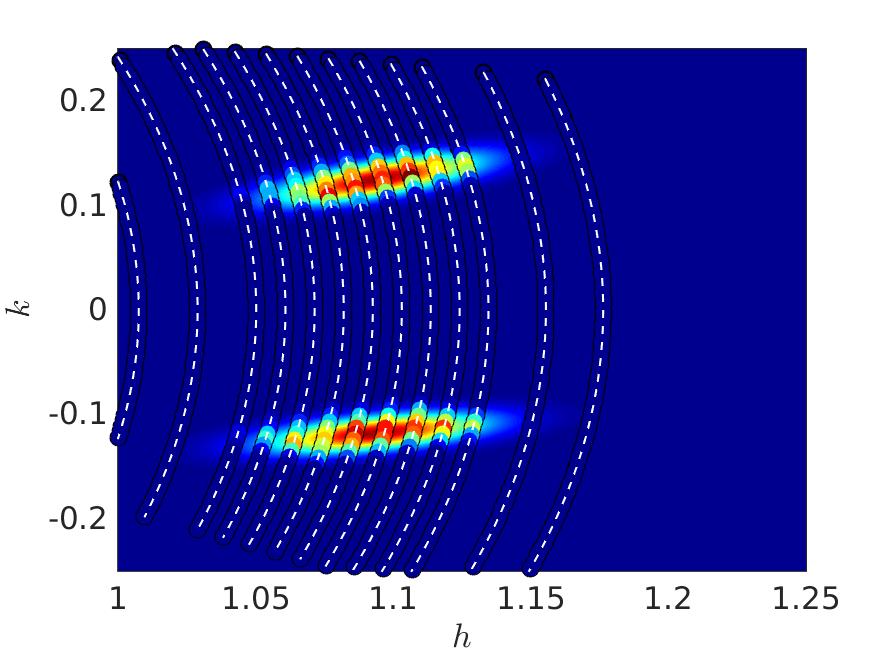}
\includegraphics[width=0.48\textwidth]{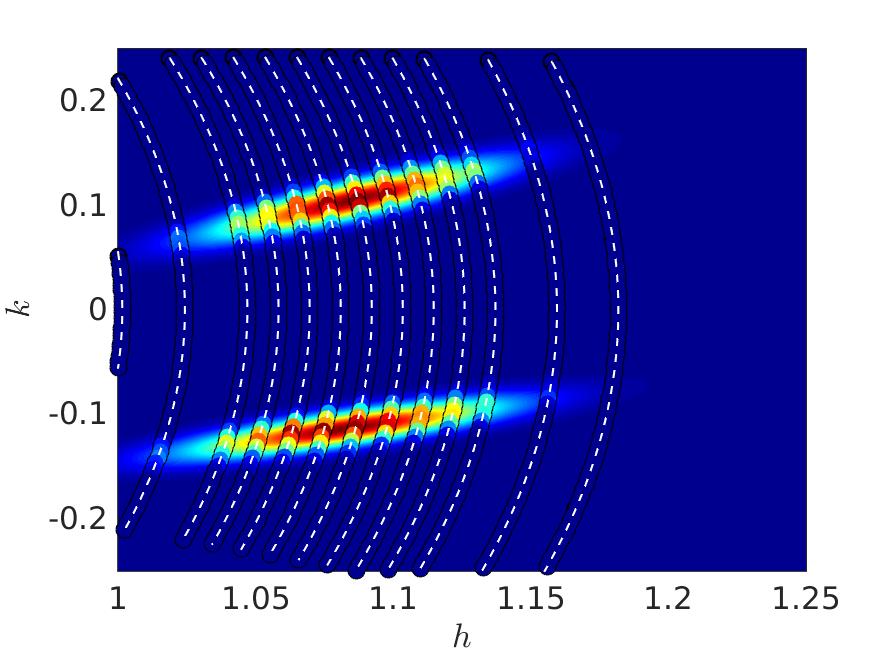} \\
\caption{Results of virtual experiments on IC peaks from LCO+O at ThALES. The colored circles with white lines through them are simulated data, with the fit shown underneath.  Top row: the simulated sample scatters at $(1+1/8,1/8,0)$. Bottom row: The simulated sample scatters at the $\QQQ$ value found in the experiments. Left column: $\Delta E=0$~meV. Right column: $\Delta E=1.5$ meV.}
\label{fig:simulation_grids}
\end{figure*}

For each set of simulations, similar scans to the ones used in the experiments were simulated. 
Grid scans, similar to the ones shown in Fig~\ref{fig:supplementary:2d_fits} were simulated at $\Delta E=0$ and 1.5~meV, see Fig.~\ref{fig:simulation_grids}, where the simulations are shown on top of fits to two-dimensional Gaussians. The tilt of the peaks in the $h,k$-plane slightly deviates from the data, although the width of the peaks is reproduced correctly. This deviation is likely caused by small inaccuracies in the description of the ThALES instrument, and does not influence our conclusions.

\begin{figure}
\centering
\includegraphics[width=0.48\textwidth]{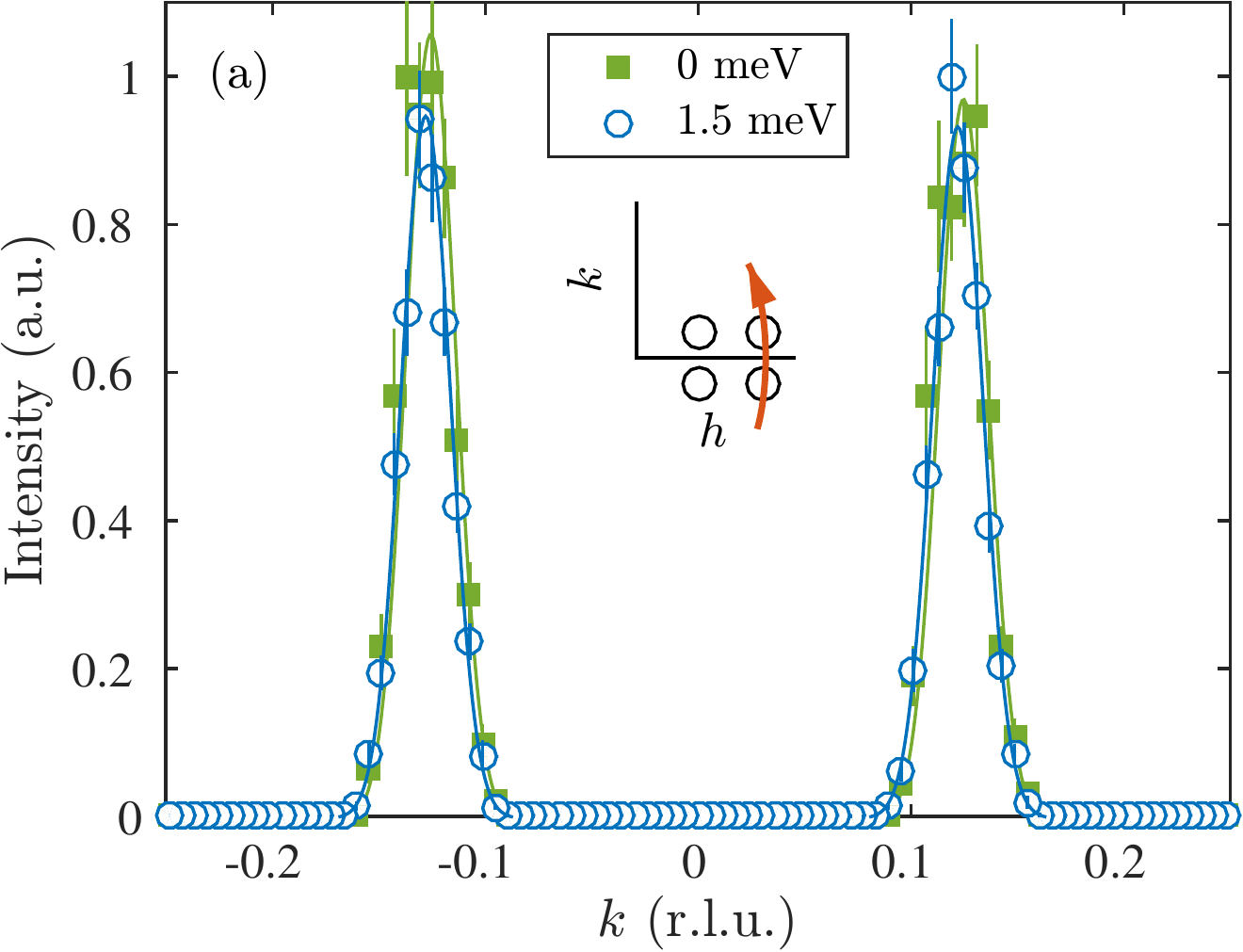}
\includegraphics[width=0.48\textwidth]{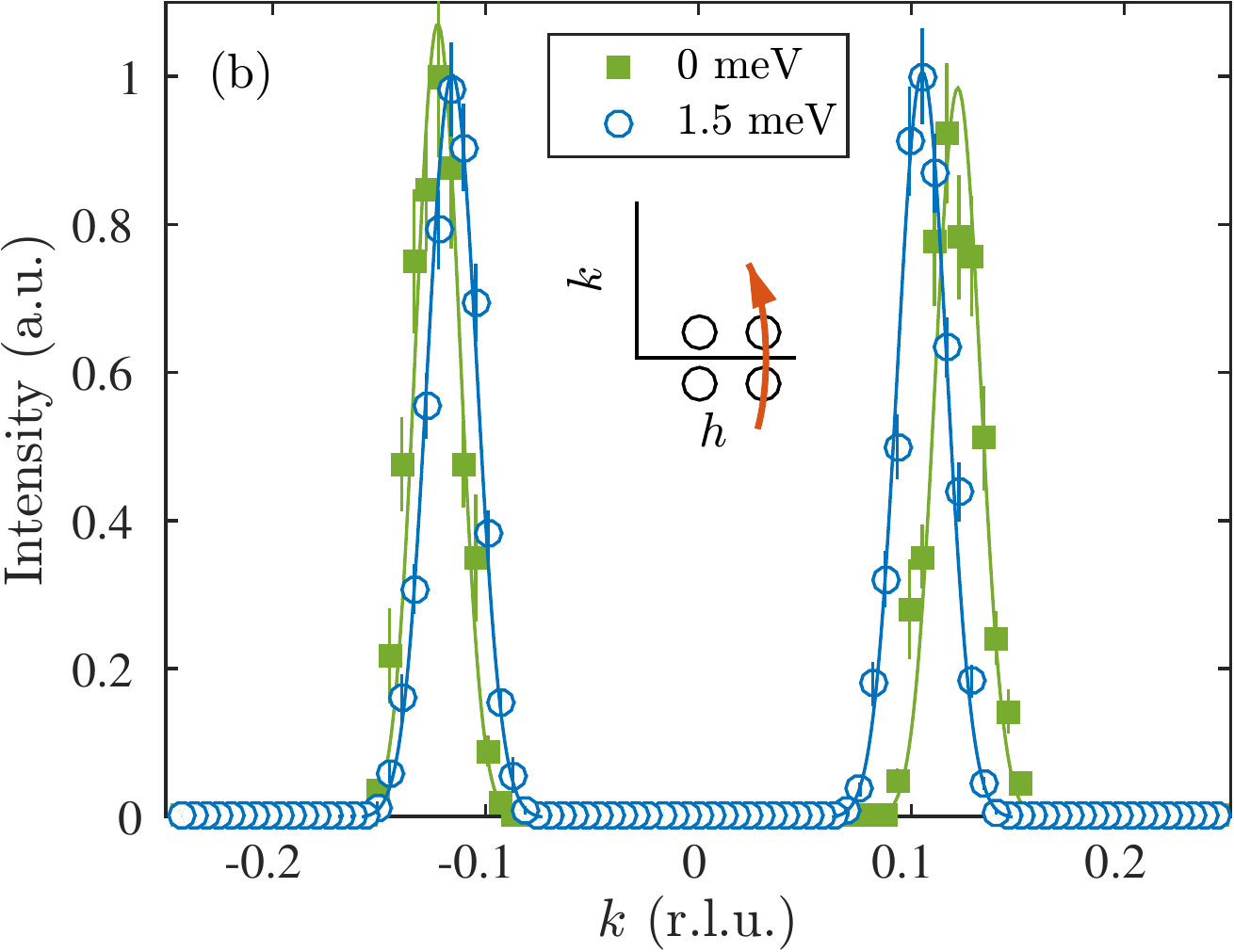}
\caption{Simulations of transverse scans on ThALES of the two IC peaks in LCO+O at $\Delta E= 0$ meV and 1.5 meV. (a) simulations with the elastic and inelastic peaks at identical positions. (b) simulations with $\delta_h= 0.0973$, $\delta_k=0.1222$ for the elastic peaks and $\delta_h=0.1036$, $\delta_k=0.1133$ for the inelastic peaks, close to the values found in the experiments.}
\label{fig:simulation_single_scans}
\end{figure}

Single scans are shown in Fig.~\ref{fig:simulation_single_scans} for the two sets of simulations. It is evident that the resolution of the instrument does not cause a shift in the distance between the incommensurate peaks. In particular, if we assume the traditional model of steeply  dispersing stripes, the simulated data do not match the actual data (Fig.~\ref{fig:simulation_single_scans} a), whereas a model with a significant difference in the peak distance between the elastic and inelastic signals agrees with the data (Fig.~\ref{fig:simulation_single_scans} b).

In Fig.~\ref{fig:delta_vs_h}, we show that the experimental observation of $\delta_k$ depending on $h$ is reproduced in the simulations. Here it is also clear that a shift in peak position between the elastic and inelastic data is needed to explain the data. There is a small effect of the resolution, causing the peaks to shift slightly towards smaller $h$ and $\delta_k$ at higher energies. There is also a small difference between the simulated and measured values of $\delta_k$ as function of $h$ for the inelastic data. Both of these effects are too small by at least an order of magnitude to explain the experimental results.

We further checked the effect of the instrument not being perfectly calibrated, so that the actual value of $k_f$ differed slightly from the set value. This was done by increasing/decreasing $E_f$ by 0.05~meV, while keeping all other parameters constant. Here, $E_f$ refers to the energy that the analyzer was aligned to select.
This did indeed cause a small shift of the peak position in $h$, but there was no change in $\delta_k$, as expected.

In total, our simulations show that the experimental observations are not caused by instrumental effects such as  resolution or misalignment.

\section{Twinning}
The stripe modulations in LCO+O are roughly along the Cu-O-Cu bonds \cite{Lee1999}, 45$^\circ$ to the orthorhombic $a$ (and $b$) axis, but parallel to the tetragonal $a$ axis as shown in Fig.~1 in the main paper. 

\begin{figure*}
    \centering
    \includegraphics[width=0.9\textwidth]{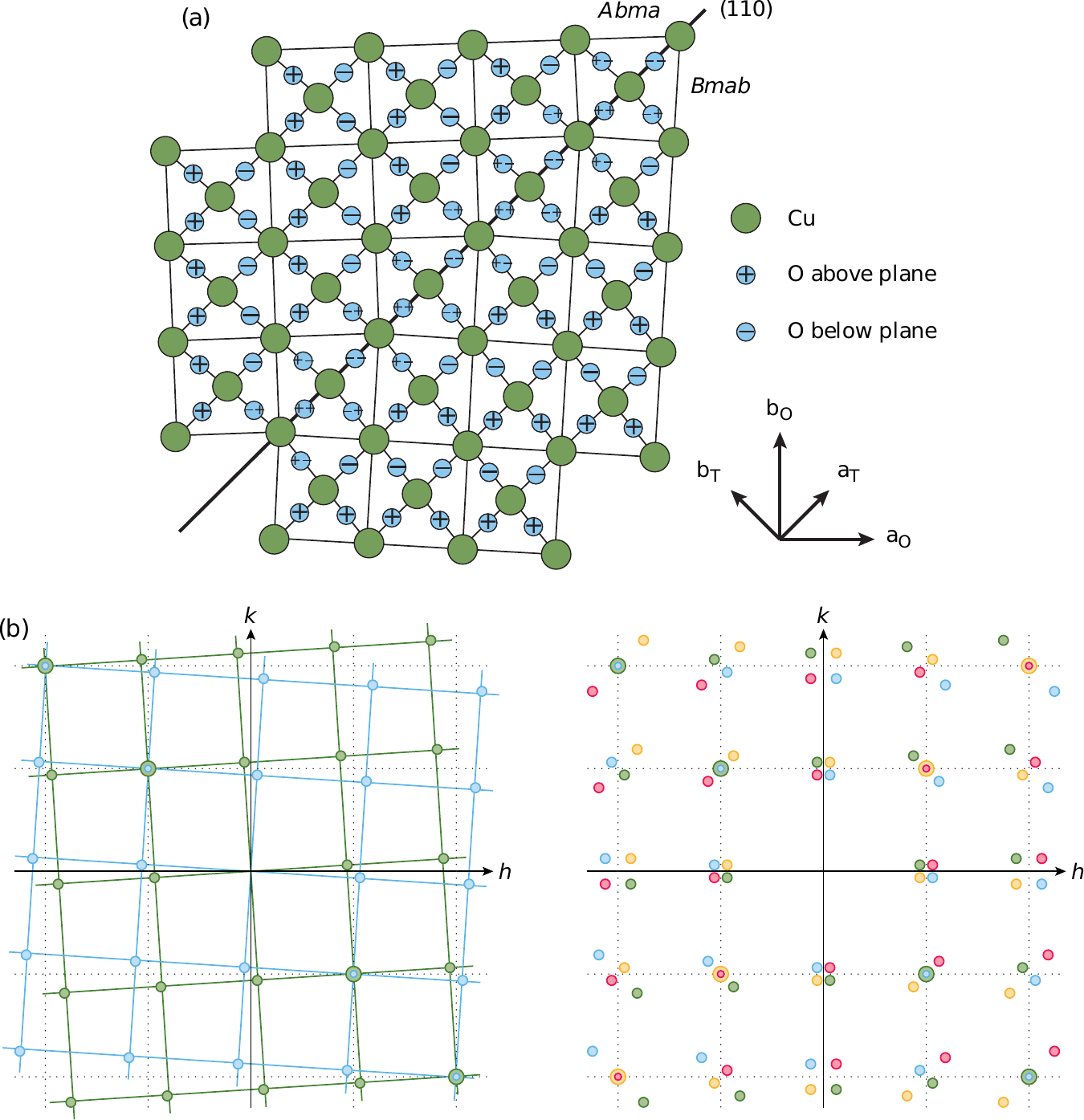}
    \caption{Illustration of twinning.
    (a) The structure of a crystal around a domain wall along (110).
    (b) The resulting reciprocal lattice from twinning along (110).
    (c) The full reciprocal lattice for twinning along (110) and (1$\bar{1}$0). Adapted from Ref.~\cite{Braden1992}\cite{piathesis}.}
    \label{fig:twinning_crystal_illustration}
\end{figure*}

Twinning occurs when cooling through the transition from a tetragonal to an orthorhombic unit cell, where $a\neq b$. In LCO+O, $a=5.33$~\AA{}, $b=5.40$~\AA{} at low temperature \cite{Lee1999}. The twinning is caused by the oxygen octahedra tilting around different axes, which slightly rotates the crystallographic axes of the different domains, see Fig.~\ref{fig:twinning_crystal_illustration}a. The results in reciprocal space is that each peak is split into two as shown in Fig.~\ref{fig:twinning_crystal_illustration}b. Similarly, a domain wall can run along the (1$\bar{1}$0) direction, creating a second set of twins, giving rise to a total of four close-lying peaks illustrated in Fig.~\ref{fig:twinning_crystal_illustration}c \cite{Braden1992}. 

\begin{figure}
    \centering
    \includegraphics[width=0.48\textwidth]{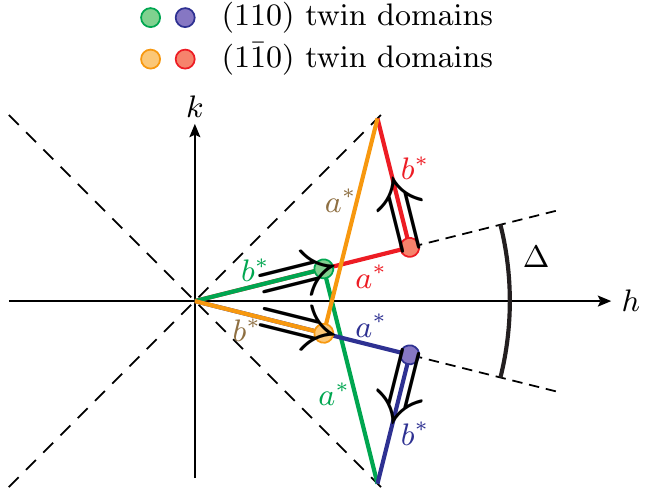}
    \caption{Illustration of the four twin peaks that appear in LCO+O around (100), with the arrows indicating the spin direction for each twin. The difference in the length of $a^*$ and $b^*$ is greatly exaggerated for clarity. }
    \label{fig:twinning_explanation}
\end{figure}

Two of the peaks originate from domains with the $a$ axis along the experimental "(100)" direction, and two with the $b$ axis along this direction, see Fig.~\ref{fig:twinning_explanation} where the direction of the spins in each domain is also shown.

Each set of two peaks with similar orientation is split by an angle 
\begin{align}
    \Delta = 90^\circ - 2 \tan^{-1}\left(\frac{a}{b}\right),
\end{align}
which in our case is around 0.7$^\circ$, or $\sim 0.01$~r.l.u. at the position of the IC peaks - too small to separate the peaks with our resolution, see Fig.~\ref{fig:supplementary:2d_fits}. The twinning can, however, easily be observed at {\em e.g.} the nuclear (200) Bragg peak.

\begin{figure}[t]
\includegraphics[width=0.22\textwidth]{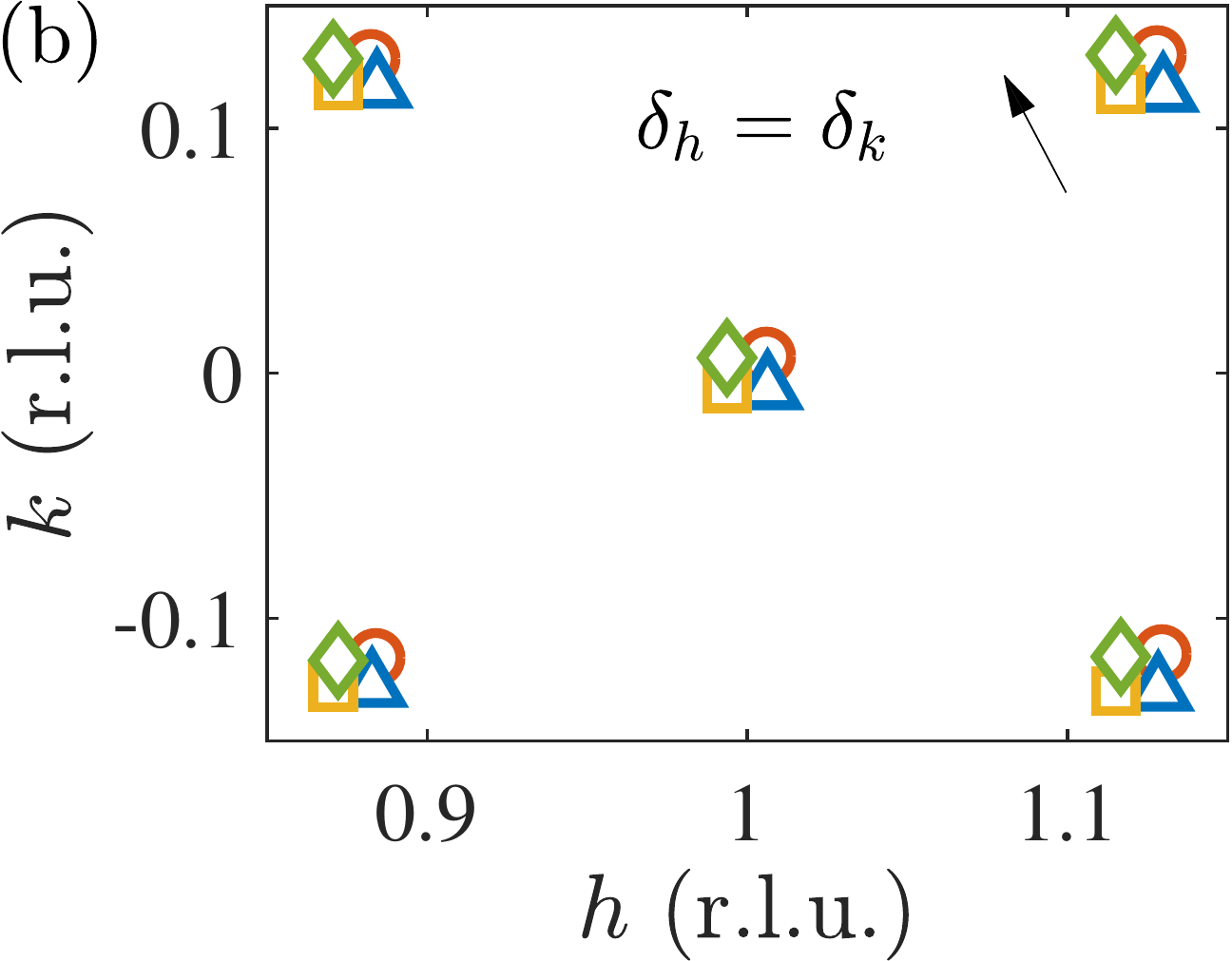}
\includegraphics[width=0.22\textwidth]{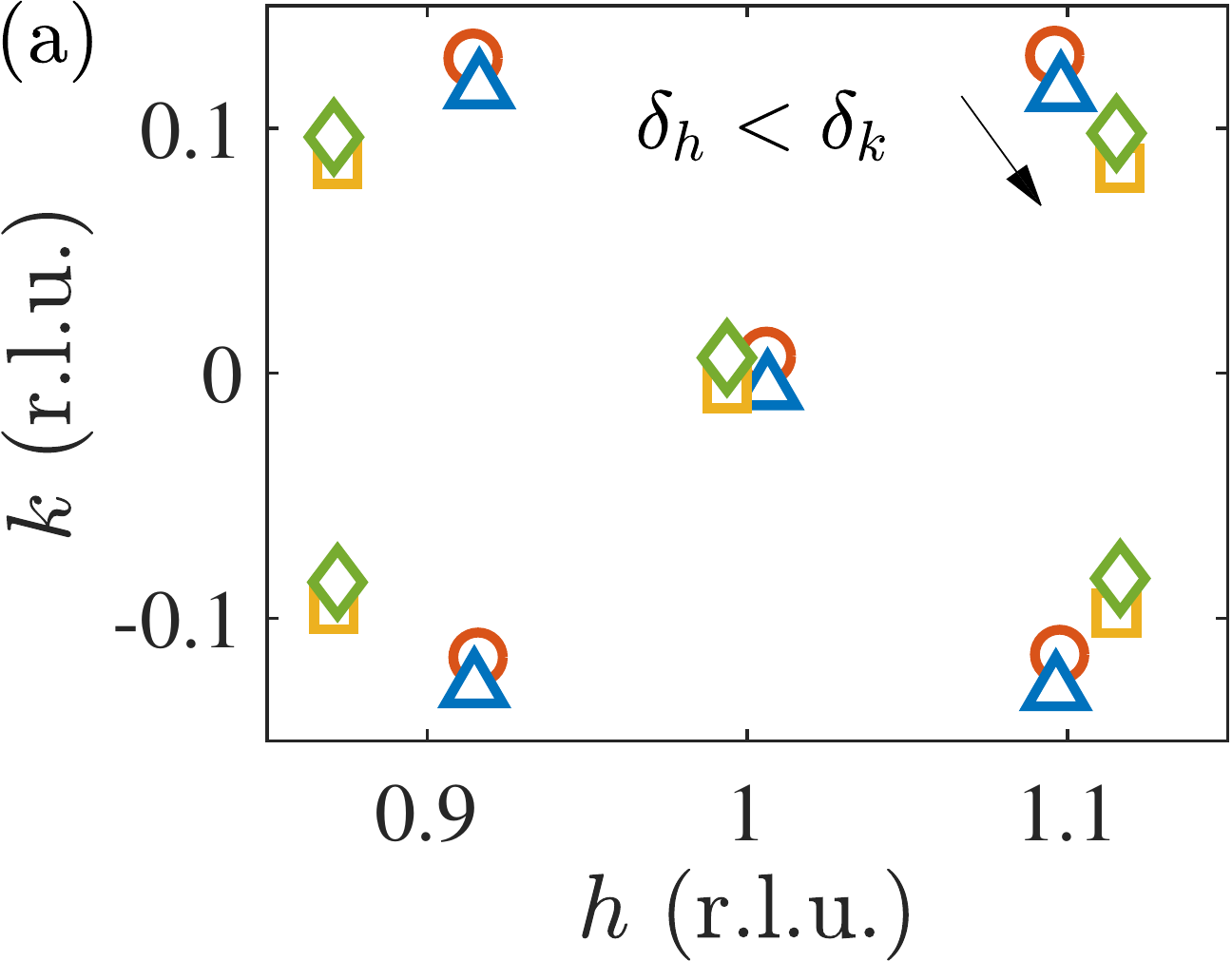}\\
\includegraphics[width=0.48\textwidth]{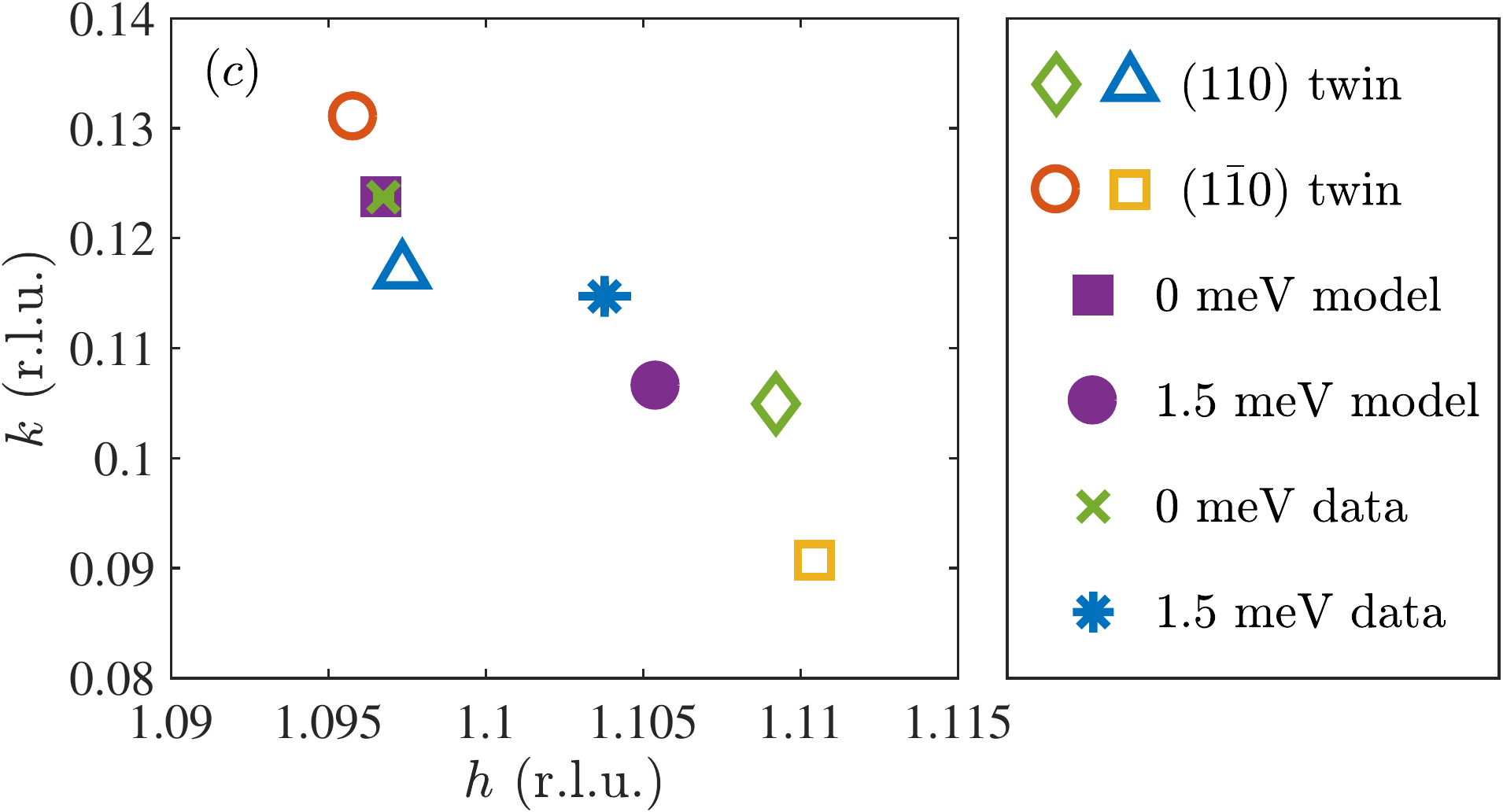} 
\caption{Illustration of reciprocal space for stripe order and twinning. 
(a) The peak positions of the stripes from the four twin domains for $\delta_h=\delta_k$, as well as the center of the peaks for each domain. 
(b) Same as (a) for $\delta_h<\delta_k$.
(c) A zoom in comparing the data with a simple model of transverse fluctuations as detailed in the text. } 
\label{fig:stripes_sketch}
\end{figure}

To estimate the effect on twinning on the observed peak positions we assume that the twin domains are equal in size and have the same stripe order with spins aligned along the local $b$ axis of the twin domain \cite{Lee1999}. The stripe order in each domain will give rise to a quartet of peaks centered around the (100), (010), ($\bar{1}$00) and (0$\bar{1}$0) positions, leading to a total of 16 peaks around the experimental (100) position, as illustrated in Fig.~\ref{fig:stripes_sketch}a. Each group of four peaks around $(1\pm \delta_h, \pm \delta_k,0)$ overlap and cannot be distinguished within the instrument resolution, and will be observed as a single peak. The position of the signal observed with neutron scattering is the average of the position of the four peaks that contribute, weighted by their relative intensity. 

In magnetic neutron scattering, the intensity is proportional to the component of $\sss$ perpendicular to $\QQQ$. In this experiment, due to twinning, we measure both $\QQQ\approx (100)$ and $\QQQ \approx (010)$. Since the spins lie along the local $(010)$ direction, there are two domains which have $\sss \perp \QQQ$ (red circle and blue triangle in Fig.~\ref{fig:stripes_sketch}), while two have $\sss \parallel \QQQ$ (green diamond and yellow square in Fig.~\ref{fig:stripes_sketch}). 
In the measurement of the elastic stripe signal, only the two former peaks  will have measurable intensity, while the two latter  will be suppressed. In this discussion we assume for simplicity that ${\QQQ \parallel (100)}$; the error in the calculated peak position from this assumption is around 1\% and thus negligible in this context.

Considering the dynamic signal and assuming the fluctuations to be transverse, each of the four peaks will have a non-zero contribution to the observed neutron scattering signal. The position of this signal will thus be different than that of the elastic signal where only two of the domains contribute.  The resulting observed shift in peak position is illustrated by the arrow in Fig.~\ref{fig:stripes_sketch}a for $\delta_h=\delta_k$ and Fig.~\ref{fig:stripes_sketch}a for $\delta_h<\delta_k$.

In Fig.~\ref{fig:stripes_sketch}c we zoom in on one of the peak positions for $\delta_h<\delta_k$, showing where the expected peaks from the four domains will be in the model described here, along with the data. We also show where the peak would be observed in a neutron scattering experiment assuming the model explained above. 
We note that in this model it is a requirement that $\delta_h<\delta_k$ as is true in our case. If $\delta_h=\delta_k$, as is usually the case in cuprates, the shift would be in the opposite direction, as evident in Fig.~\ref{fig:stripes_sketch}a.

\end{document}